# An origin for short γ-ray bursts unassociated with current star formation

S. D. Barthelmy[1], G. Chincarini[2,3], D. N. Burrows[4], N. Gehrels[1], S. Covino[2], A, Moretti[2], P. Romano[2], P. T. O'Brien[5], C. L. Sarazin[6], C. Kouveliotou[7], M. Goad[5], S. Vaughan[5], G. Tagliaferri[2], B. Zhang[8], L. A. Antonelli[9], S. Campana[2], J. R. Cummings[1,10], P. D'Avanzo[2,11], M. B. Davies[12], P. Giommi[13], D. Grupe[4], Y. Kaneko[14], J. A. Kennea[4], A. King[5], S. Kobayashi[4], A. Melandri[9], P. Meszaros[4,15], J. A. Nousek[4], S. Patel[14], T. Sakamoto[1,10] & R. A. M. J. Wijers[16]

[1]*NASA/Goddard Space Flight Center Greenbelt, Maryland 20771, USA.* [2]*INAF—Osservatorio Astronomico di Brera, Via Bianchi 46, I-23807 Merate, Italy.* [3]*Universita degli studi di Milano Bicocca, Piazza delle Scienze 3, I-20126 Milano, Italy.* [4]*Department of Astronomy and Astrophysics, Penn State University, University Park, Pennsylvania 16802, USA.* [5]*Department of Physics and Astronomy, University of Leicester, Leicester LE1 7RH, UK.* [6]*Department of Astronomy, University of Virginia, Charlottesville, Virginia 22903-0818, USA.* [7]*NASA/Marshall Space Flight Center, NSSTC, XD-12, 320 Sparkman Drive, Huntsville, Alabama 35805, USA.* [8]*Department of Physics, University of Nevada, Las Vegas, Las Vegas, Nevada 89154-4002, USA.* [9]*Osservatorio Astronomico di Roma, Via di Frascati, 33 00040 Monte Porzio Catone, Italy.* [10]*National Research Council, 2101 Constitution Avenue NW, Washington, DC 20418, USA.* [11]*Universita degli Studi dell'Insubria, Dipartimento di Fisica e Matematica, via Valleggio 11, 22100 Como, Italy.* [12]*Lund Observatory, Box 43, SE-221 00 Lund, Sweden.* [13]*ASI Science Data Center, Via Galileo Galilei, I-00044 Frascati, Italy.* [14]*Universities Space Research Association, NSSTC, XD-12, 320 Sparkman Drive, Huntsville, Alabama 35805, USA.* [15]*Department of Physics, Penn State University, University Park, Pennsylvania 16802, USA.* [16]*Astronomical Institute 'Anton Pannekoek', University of Amsterdam, Kruislaan 403, 1098 SJ Amsterdam, The Netherlands.*

**Two short (<2 s) γ-ray bursts (GRBs) have recently been localized[1–4] and fading afterglow counterparts detected[2–4]. The combination of these two results left unclear the nature of the host galaxies of the bursts, because one was a star-forming dwarf, while the other was probably an elliptical galaxy. Here we report the X-ray localization of a short burst (GRB 050724) with unusual γ-ray and X-ray properties. The X-ray afterglow lies off the centre of an elliptical galaxy at a redshift of $z=0.258$ (ref. 5), coincident with the position determined by ground-based optical and radio observations[6–8]. The low level of star formation typical for elliptical galaxies makes it unlikely that the burst originated in a supernova explosion. A supernova origin was also ruled out for GRB 050709 (ref. 3), even though that burst took place in a galaxy with current star formation. The isotropic energy for the short bursts is 2–3 orders of magnitude lower than that for the long**

bursts. Our results therefore suggest that an alternative source of bursts—the coalescence of binary systems of neutron stars or a neutron star-black hole pair—are the progenitors of short bursts.

On 2005 July 24 at 12:34:09.32 UT (Universal Time), the BAT instrument on the *Swift* observatory[9] triggered and located GRB 050724 (ref. 6). The lightcurve for this event (Fig. 1) is somewhat different from the typical short GRB population as originally defined by the Burst and Transient Source Experiment (BATSE)[10]. Similar to GRB 050709 (ref. 11), the event consists of two parts that are spectrally very different. The main portion has a duration of 3.0±1.0 s and is dominated by a hard spike lasting only 0.25 s. Following the main peak at time $T$, there is a second smaller peak at $T$+1.1 s plus long-lasting faint emission out to $T$+140 s, which is all in the 15–25 keV band. The faint tail images to the same location on the sky as the initial hard spike and is therefore definitely part of the initial event. The fluence in the faint tail is only 10% of that in the main peak.

The hardness ratio (50–100 keV to 25–50 keV) of the first peak is 0.91±0.12, which is on the low end of the distribution for short GRBs as detected with the BATSE instrument. The hardness ratio is 0.55±0.23 for the second peak at $T$+1.1 s and 0.31±0.10 for the long-lasting emission out to $T$+140 s. To determine whether this GRB would indeed have been classified as short in the BATSE sample, we performed a series of 100 simulations each, using the BATSE large-area detector response matrices for three different angles of incidence, for all main constituents of GRB 050724. We found that in its typical trigger energy range (50–300 keV), BATSE would have triggered on the main hard peak and would definitely have detected the second pulse at $T$+1.1 s; however, the softer pulse starting at $T$+30 s would have been detected at best at the 0.3$\sigma$ level and, thus, it would have not been accounted as part of the GRB. As a final test we passed the entire BAT lightcurve collected between 50–350 keV through the BATSE $T_{90}$ algorithm, obtaining $T_{90}$=1.31±0.53 s, entirely consistent with the short GRB population (but does not rule out a long-burst origin). Therefore, the long-lasting emission of this burst is probably a newly detected component of short bursts detected by BAT with its 15 keV threshold and not seen by BATSE with its >25 keV threshold (although possibly hinted at in the sum of multiple short BATSE GRBs[12,13]).

The Swift X-Ray Telescope (XRT) began taking data 74 s after the BAT trigger[6], and located a bright, fading X-ray source, identified as the X-ray afterglow[14,15]. Ground-based observatories detected radio[7] and optical[16,17] afterglows within the XRT error circle and we obtained a high-precision X-ray position[8] with our Chandra X-ray Observatory target of opportunity observation. The positions determined by the various instruments are given in Table 1 and shown in Fig. 2. The radio, optical, and X-ray positions are all coincident.

The fading afterglow of GRB 050724 was located[18] approximately 1" south of the centre of (but within) a bright elliptical galaxy at a redshift of $z=0.258$ (ref. 5). The projected offset from the centre of the galaxy corresponds to ~4 kpc (see Fig. 2). The host galaxy has an apparent magnitude of $K=15.3$ within a 3" radius aperture[17], which corresponds to a luminosity of $2\times10^{11}L_O\approx1.7L^*$ (where $L_O$ and $L^*$ are the luminosities of the Sun and a typical galaxy, respectively). The total luminosity is probably slightly larger. Galaxies this luminous are relatively uncommon, with a comoving number density[19] of only $3\times10^{-4}$ Mpc$^{-3}$. The probability that a GRB would occur randomly within 1" of the centre of a galaxy this luminous at a redshift of $z<0.3$ is only ~$10^{-5}$. The small probability of an accidental superposition and lack of any other potential GRB host galaxies within the GRB error circle make the identification of this elliptical as the host galaxy quite secure.

The host galaxy's red colour[20], spectrum and visual appearance are all consistent with a luminous elliptical galaxy and very similar to the properties of the host galaxy for the first localized short burst, GRB 050509B (ref. 1). One difference is that the host of GRB 050509B was located in a moderately rich cluster of galaxies, while the optical and X-ray observations of GRB 050724 suggest that this host elliptical is located in a lower-density region. The spectrum of the host shows no emission lines[18] or evidence for recent star formation, and is consistent with a population of very old stars. This is true of most large elliptical galaxies in the present-day Universe, including the host galaxy of GRB 050509B. The elliptical hosts of these two short GRBs are very different from those for long bursts, which are typically sub-luminous, blue galaxies with strong star formation[21].

Thus the properties of these two short GRB hosts suggest that the parent populations and consequently the mechanisms for short and long GRBs are different in significant ways. Their non-star-forming elliptical hosts indicate that short GRBs could not have resulted from any mechanism involving massive star core collapse[22] or recent star formation (for example, a young magnetar giant flare[23,24]). As we previously noted[1], large elliptical galaxies are very advantageous sites for old, compact binary star systems, and thus good locations for neutron star–neutron star or neutron star–black hole mergers. Luminous elliptical galaxies are known to contain large populations of low-mass X-ray binaries containing neutron stars or black holes, and have large numbers of globular clusters within which compact binary stars can be formed dynamically with a much higher efficiency than in the field. Note, however, that mergers of compact objects are also expected to occur with a significant rate in star-forming galaxies; even if such mergers are the mechanism behind all short GRBs, one would not expect them all to occur in elliptical galaxies. In fact, the second short GRB with fine localization (GRB 050709)[2–4] was in a star-forming galaxy at $z$=0.16 and may be such a case.

Taking into account the host distance, we compare the energetics of short and long GRBs. The fluence in the first 3 s of emission is $6\times10^{-7}$ erg cm$^{-2}$ in the 15–350 keV range, which translates roughly to a total 10 keV–1 MeV γ-ray fluence of ~$10^{-6}$ erg cm$^{-2}$. The fluences in the 30 to 200 s soft γ-ray peak and the X-ray afterglow are comparable at $7\times10^{-7}$ erg cm$^{-2}$ and ~$10^{-6}$ erg cm$^{-2}$, respectively. These fluences are similar to those seen by BAT and other γ-ray detectors for long bursts. However, at a redshift of $z$=0.285, the total energy output of the prompt plus soft peak is ~$3\times10^{50}$ erg, which is significantly less than the $10^{52}$–$10^{54}$ erg isotropic-equivalent energy output of the long GRB class[25].

The early XRT lightcurve initially shows a steep decay with a slope of _2. This component is connected to the soft γ-ray component around ~80 s identified in the extrapolated BAT lightcurve, which clearly overlaps and joins the early XRT measurements (Fig. 3). This flare-like event is followed by a very rapid decay after ~100 s (with index <–7), interrupted by a second, less-energetic flare around 200–300 s. The curve flattens at 700 s and has a roughly constant decay index all the way through the Chandra points to $2\times10^5$ s. A third significant flare starting at ~$2\times10^4$ s is superposed on this decay component, with the total energy even smaller (by a factor of ~3) than the second one. The steep decays following all three flares are too steep to be interpreted as the afterglow emission from a forward shock, but could be consistent with the high latitude emission ($\theta$>>1/$\Gamma$) from a fireball that suddenly stops radiating or goes off in an extremely low-density medium (naked GRBs)[26]. This is pertinent for the late internal shock scenario as invoked to interpret the X-ray flares in long GRBs[27]. This interpretation requires that the central engine remains active up to at least ~200 s (which is supported by the flaring event in the BAT-extrapolated X-ray lightcurve and the wiggles in the initial XRT lightcurve).

The current neutron star–neutron star merger models[28,29] predict energy injection times much shorter than the >200 s time seen for GRB 050724. Black hole–neutron star mergers are more promising[30], because they allow for partial disruption of the neutron star and gradual accretion as the higher angular momentum material decays through gravitational radiation, but even these models cannot extend the emission beyond a few tens of seconds. With the current Swift detection rate of one short burst per 2–3 months, the sample will quickly increase and will provide answers about how typical extended emission is for short burst.

**Acknowledgements** We acknowledge support from ASI, NASA and PPARC, and acknowledge benefits from collaboration within the EU FP5 Research Training Network 'γ-Ray Bursts: An Enigma and a Tool'.

Correspondence and requests for materials should be addressed to S.D.B. ([scott@lheamail.gsfc.nasa.gov](scott@lheamail.gsfc.nasa.gov)).

**Table 1** Position determinations for GRB 050724. All the positions are consistent with each other to within the errors quoted for each. See Fig 2.

| Observatory | RA (J2000) | Dec. (J2000) | Error circle radius* | Notes | Ref. |
|---|---|---|---|---|---|
| Swift/BAT | 16 h 24 min 43 s | _27° 31_ 30__ | 3′ | 1_ from Chandra position | 6 |
| Swift/XRT | 16 h 24 min 44.41 s | _27° 32_ 28.4__ | 6__ | Corrected astrometry relative to position in GCN Circular 3678 | 15 |
| VLT | 16 h 24 min 44.37 s | _27° 32_ 27__ | 0.5__ | | |
| VLA | 16 h 24 min 44.37 s | _27° 32_ 27.5__ | 0.2__ | One-sigma error | 7 |
| Chandra/ACIS | 16 h 24 min 44.36 s | _27° 32_ 27.5__ | 0.5__ | | 8 |

*90% confidence limit except for VLA. VLT, Very Large Telescope. VLA, Very Large Array. RA, right ascension; Dec., declination.

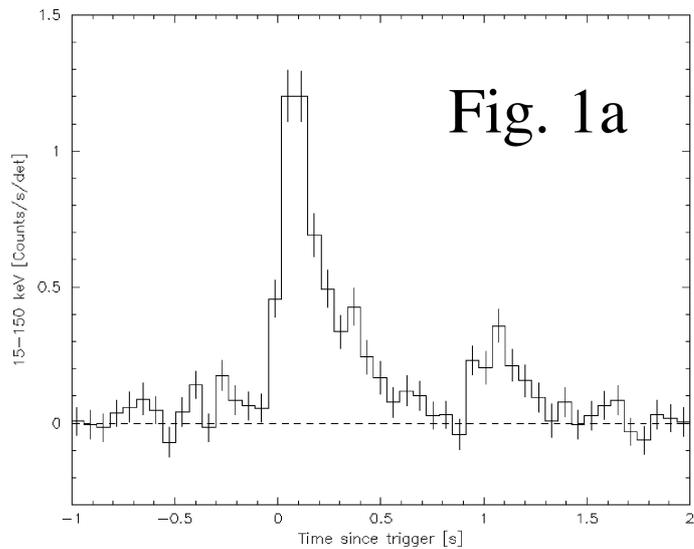

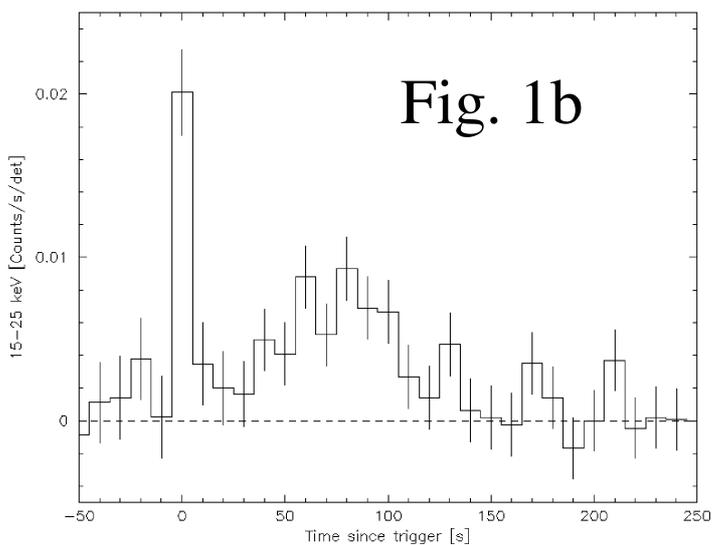

**Figure 1 BAT lightcurves for GRB 050724 showing the short duration of this GRB and the long softer emission. a**, The prompt emission in the 15–150 keV energy band with a short-duration main spike of 0.25 s. $T_{90}$ is 3.0±1.0 s ($T_{90}$ is the time during which 90% of the GRB photons are emitted[10]; the fluence is $(3.9\pm1.0)\times10^{-7}$ erg cm$^{-2}$ and the peak flux is 3.5±0.3 photons cm$^{-2}$ s$^{-1}$ (15–150 keV, 90% confidence level). **b**, Soft emission in the 15–25 keV energy band lasting >100 s (peak flux is ~$2\times10^{-9}$ erg cm$^{-2}$ s$^{-1}$). The error bars in both panels are one-sigma standard deviation. The BAT energy spectrum in the prompt portion ($T\_0.03$ to $T+0.29$ s; where $T$ equals BAT trigger time of 12:34:09.32 UT) is well fitted with a simple power-law model of photon index 1.38±0.13 and normalization at 50 keV of 0.063±0.005 photons cm$^{-2}$ s$^{-1}$ keV$^{-1}$ (15–150 keV, 90% confidence level).

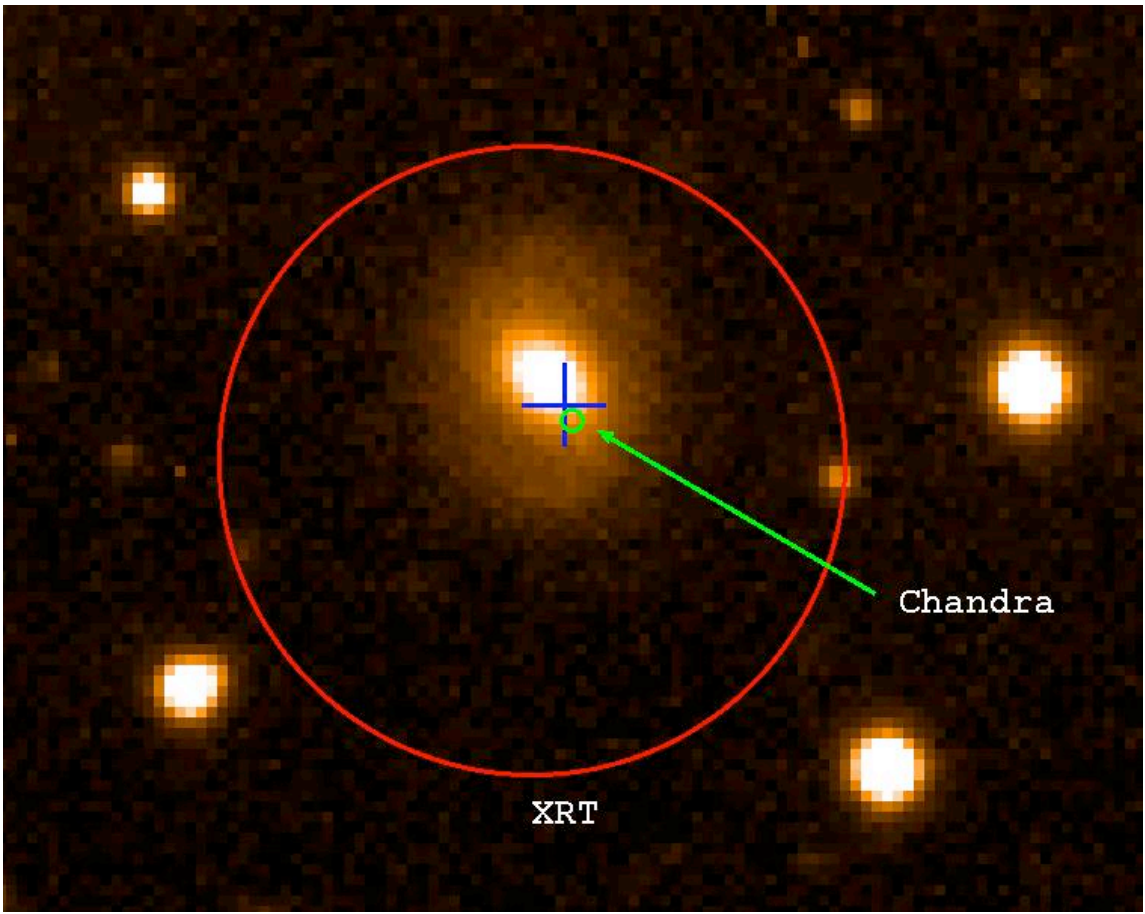

**Figure 2 VLT optical image[17] showing the association of GRB 050724 with the galaxy.** The blue cross is the position of the optical transient[16,17]. The XRT (red circle) and Chandra (green circle) burst positions are superimposed on a bright red galaxy at redshift $z$=0.258 (ref. 5), implying a low-redshift elliptical galaxy as the host. The XRT position has been further revised from the position of ref. 15 by astrometric comparison with objects in the field. The projected offset from the centre of the galaxy corresponds to ~4 kpc assuming the standard cosmology with $H_0$=71 km s$^{-1}$ Mpc$^{-1}$ and ($\Omega_M$, $\Omega_\Lambda$)=(0.27, 0.73).

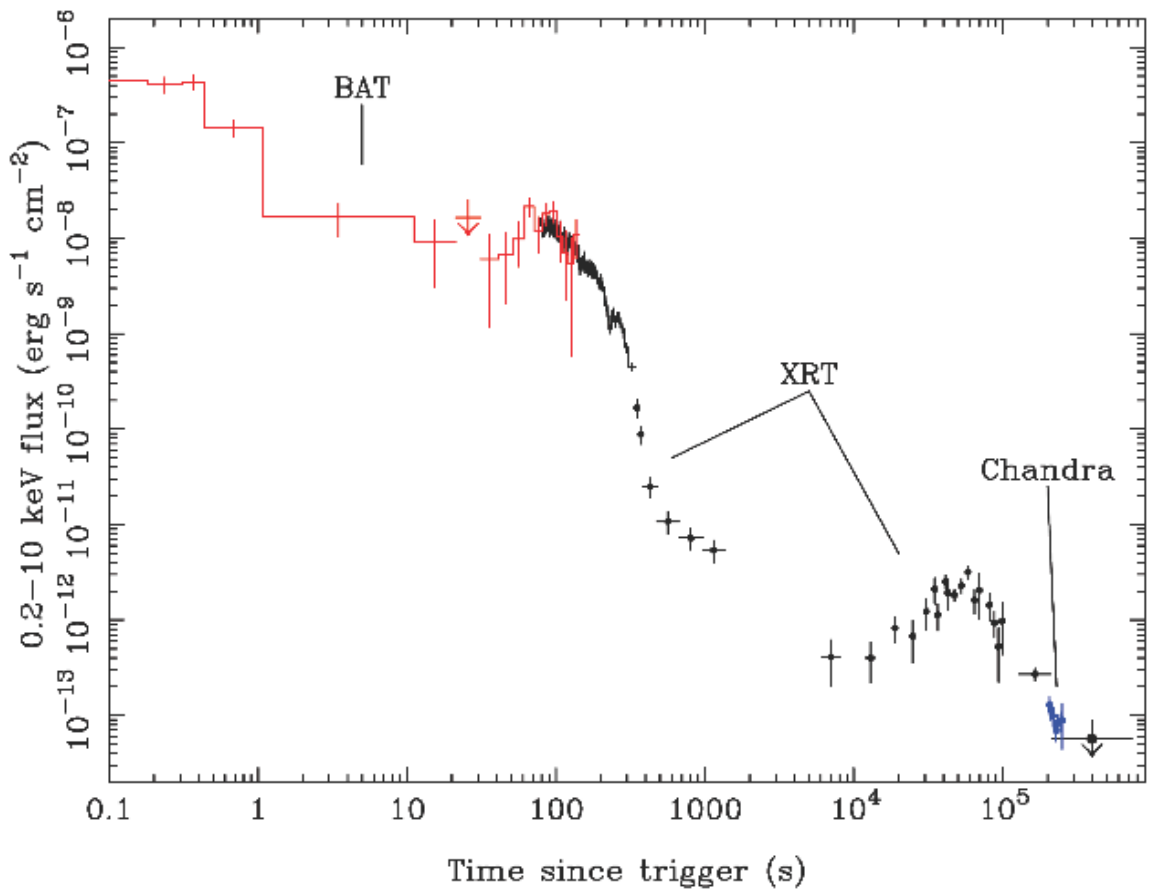

**Figure 3 The smooth transition of the GRB phase into the X-ray phase.** This shows the total X-ray afterglow lightcurve for GRB 050724 by combining data from BAT, XRT and Chandra. The prompt hard phase detected by BAT decays into the soft γ emission, and is followed by flaring (~$T$+80 s) and then gradual fall-off into the X-ray afterglow measured by XRT. The BAT points are binned with variable time intervals such that the signal-to-noise ratio is >5 and Δ$T$<10 s. Using the Band model to simultaneously fit the BAT spectra in the 15–150 keV band in the $T$+(50–150) s time interval with the XRT spectra in the 0.2–10 keV band in the $T$+(79–150) s interval (yielding $E_{break}$=30 keV), the BAT flux points (15–25 keV band) were scaled down into the X-ray band (unabsorbed 0.2–10 keV). The Band model was required because power-law fits to the BAT and XRT separately yielded $\Gamma$ values of 2.5 and 1.7, respectively: that is, there was a spectral break between the two instruments. The XRT fluxes are unabsorbed values derived using best-fit XRT photon indices of $\Gamma$=1.9 (for the 'window timing' mode data before 341 s) and $\Gamma$=1.5 (for the 'photon counting' mode data after 342.9 s) and using the best-fit absorbing column density of $N_H$=5.9×10$^{21}$ cm$^{-2}$. All error bars are one-sigma standard deviation. The Chandra points are unabsorbed fluxes using the XRT-measured value for $N_H$ and the Chandra-measured best-fit photon index of $\Gamma$=1.8.